\documentclass[doublecol]{epl2} 
\usepackage{amsmath}
\title{Universality in the jamming limit for elongated hard particles in one dimension}
\shorttitle{Universality in the jamming limit for elongated hard particles in 1D} 
\author{Yacov Kantor\inst{1} \and Mehran Kardar\inst{2}}
\shortauthor{Yacov Kantor and Mehran Kardar}
\institute{                    
  \inst{1} Raymond and Beverly Sackler School of Physics and
 Astronomy, Tel Aviv University, Tel Aviv 69978, Israel\\
  \inst{2} Department of Physics, Massachusetts Institute of
 Technology, Cambridge, Massachusetts 02139, USA
}

\abstract{
We study thermodynamics properties of a one dimensional gas of hard elongated 
particles. The particle centers are restricted to a line, while they can rotate 
in two-dimensional space. Correlations between orientations of the objects are 
studied (by transfer matrix method) as a function of density and aspect ratio.
The behavior in the extreme high-density (jamming) limit is described by a few 
universality classes depending on the object's shape. In particular, there is 
a diverging correlation length when the contact point  of  adjacent 
objects is far from the line along which their centers move, as for 
needles and rectangles.}

\begin{document}
\maketitle

One-dimensional (1D) collections of classical particles are usually simple to study,
yet can offer valuable insights into collective properties of higher dimensional systems.
In particular, thermodynamic properties of a gas of particles confined to a line,
and interacting with a short-range potential can be solved exactly~\cite{takanashi}.
For example, the solvable 1D system of ``hard spheres," sometimes referred to as the
{\em Tonks  gas}~\cite{tonks}, served as an initial step in the study of two-
and three-dimensional systems of hard disks/spheres, and more recently was invoked
in connection with proteins on DNA~\cite{chou}.

Usage of {\em hard} particles, whose interaction energy is either 0 or $\infty$,
is yet another useful simplification that highlights the geometric/entropic 
features of interacting systems \cite{gast}. 
The absence of an energy scale renders the behavior independent of temperature,
emphasizing the variations with shape and density.
Indeed, numerical studies of hard potentials date back to the origins of
the Metropolis Monte Carlo method~\cite{origMetro}.  
More recently, dynamic properties of hard particles have been studied in connection
with granular flows~\cite{edwards} and jamming transitions~\cite{jamming}.
Critical behavior and diverging correlation lengths at the so-called J-point (for jamming) have
been studied from several perspectives\cite{bouchaud,chayes,nelson,parisi}.

\begin{figure}
\onefigure[width=7.5cm]{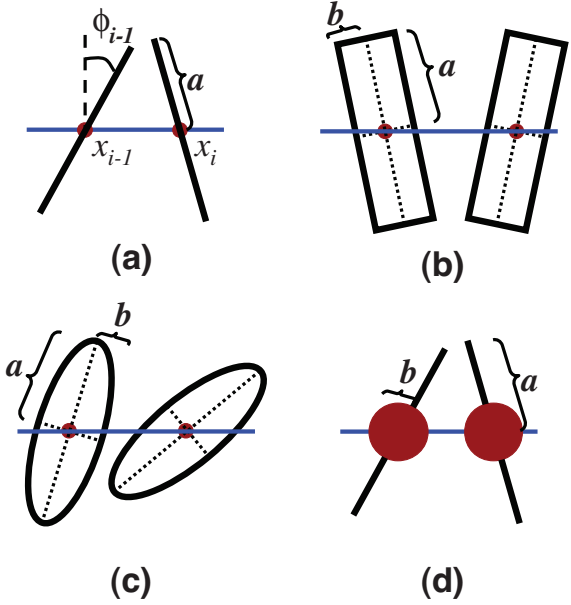}
\caption{\label{fig:objects} (Color online) Some elongated objects considered
in this work: (a)~needles, (b)~rectangles, (c)~ellipses, and (d)~metronomes. 
Centers of these two-dimensional objects are placed on a straight line. 
The degrees of freedom for each particle are the position of its center 
$x_i$ (constrained to 1D), and the angle $\phi_i$ between its major axis and 
the normal to the line,  as depicted in (a). The longer and shorter semi-axes 
of the particles  are denoted $a$  and $b$, respectively. 
All particles in a particular  ensemble are identical. 
Needles can be viewed as the $b\to 0$ limit of either  ellipses or rectangles.
 }
\end{figure}

Here, we demonstrate that diverging correlation lengths may also arise in 
the high density (jamming) limit of elongated hard particles in one-dimension 
(such as depicted in Fig.~\ref{fig:objects}). Non-spherically symmetric hard particles~\cite{sa,sb,sc,sca,scb,scc,scd,sce,scf,fthda,fthdb,fthdc,frenkel_thinrod}
provide important perspectives into the phase behavior of liquid crystals. 
Indeed, following an approach outlined in Ref.~\cite{casey},
Lebowitz {\it et al.}~\cite{lebowitz} studied this very system of elongated objects
moving on a line and rotating in the plane.
They showed that in the limit of high density, the 1D pressure acquires a simple form,
dependent on the curvatures of the object at their point of contact.
We build upon these results, focusing on the orientational fluctuations and their correlations.
We show that, depending on the shapes of the objects, there are a few generic universality 
classes in the jamming limit, some characterized by diverging correlation lengths.

%%%%%%%%%%%%%%%%%%%%%%%%%%%%%%%%%%%%%%%%%%%%%%%%%%%%%%%%%%
%%%%%%%%%%%%%%%%%%%%%%%%%%%%%%%%%%%%%%%%%%%%%%%%%%%%%%%%%%

As indicated in Fig.~\ref{fig:objects}, the elongated particles considered in our work
are characterized by the positions $\{x_i\}$ of their centers of mass,
and their orientations $\{\phi_i\}$ (measured with respect to the normal to the line).
We consider symmetric objects that are invariant under a change of $\phi_i$ by $\pi$, 
and thus restrict $-\pi/2\le\phi_i<\pi/2$. 
(There are obvious generalizations to three dimensional objects, with more angular
degrees of freedom, which shall not be pursued here.)
The elongated objects are typically described by long and short  semi-axis lengths 
$a$ and $b$ ($a>b$). 
The needle can be viewed as either an ellipse or a rectangle with $b=0$. 
The system consists of $N$ identical particles, and without loss of generality
we shall assume that  their order along the line remains unchanged, i.e. $x_i< x_{i+1}$. 

The particles are not allowed to intersect but do not interact otherwise.  
The minimal distance $d_{i-1,i}=d(\phi_{i-1},\phi_i)$  between two adjacent 
particles depends on their orientations $\phi_{i-1}$ and $\phi_i$, and the size 
and shape of the objects. The hard particle constraint is equivalent to the 
interaction
\begin{equation}\label{eq:V}
V\left(x_{i-1},\phi_{i-1};x_i,\phi_i\right)=
\begin{cases}
0,  & \tx{ if  } x_i-x_{i-1}>d_{i-1,i}, 
\\
\infty, & \tx{ otherwise.}
\end{cases}
\end{equation}

The standard method for dealing with potentials that depend only on the
nearest neighbor separations is to consider an ensemble of fixed pressure $p$,
and evaluate the Gibbs partition function \cite{pathria,kardar} at temperature $T$:
\begin{eqnarray}\label{eq:Zdef}
{\cal Z}&=&\int\prod_{i=1}^N\upd x_i\prod_{i=1}^N\upd \phi_i
{\rm e}^{-\beta\sum_{i=1}^NV\left(x_i-x_{i-1};\phi_{i-1},\phi_i\right)-\beta p x_N}\nonumber\\
&=&\int\prod_{i=1}^N\upd\phi_i\prod_{i=1}^N\left[\int_0^\infty \upd u_ie^{-\beta pu_i-\beta V\left(u_i;\phi_{i-1},\phi_i\right)}\right]\nonumber\\
&=&(\beta p)^{-N}\int\prod_{i=1}^N\left(\upd\phi_i{\rm e}^{-\beta p d(\phi_{i-1},\phi_i)}\right)\,,
\end{eqnarray}
where $\beta=1/k_BT$, and $k_B$ is the Boltzmann constant. In the second line 
of Eq.~\eqref{eq:Zdef}, we have changed variables to the separations 
$u_i=x_i-x_{i-1}$, setting the boundary conditions $x_0=\phi_0=0$ for convenience.
Since the potential $V$ is zero or infinity, the integration over each $u_i$ is easily
performed, leading to the final result, which only depends on the combination
$\beta p$.

As pressure is increased the particles are jammed together, and forced to align
to minimize their separations. The thermodynamic properties of the system in this limit
can be obtained by evaluating the Gibbs partition function in Eq.~\eqref{eq:Zdef} by
a saddle point expansion around $\{\phi_i=0\}$. The expansion of $d(\phi,\phi')$ depends
on the shape of the objects, and for the cases depicted in Fig.~\ref{fig:objects} we have:
\begin{itemize}
\item{For rectangles 
\begin{equation}\label{eq:d_rectangle}
d(\phi,\phi')=2b+\frac{b}{2}(\phi^2+\phi'^2)+a\left|\phi-\phi'\right|+\cdots\,.
\end{equation}
The first two terms reflect the change in the cross-section of each rectangle as it tilts.
The final term reflects the advantage of aligning neighboring rectangles.
}
\item{Needles are obtained as a limit of the above expression when $b=0$. The full
expression, including higher order terms is \cite{kkelast}
\begin{equation}\label{eq:d_needle}
d(\phi,\phi')=a\frac{\sin\left|\phi-\phi'\right|}{ \max[\cos(\phi),\cos(\phi')]}.
\end{equation}
}
\item{The minimal distance between two ellipses can be obtained by using the expressions
in Ref.~\cite{vieillard}. For small tilts from vertical, we find
\begin{eqnarray}\label{eq:d_ellipse}
d&\approx & 2b+\frac{b}{2}\left(1-\frac{b^2}{ a^2}\right)(\phi^2+\phi'^2)
\nonumber \\
 &+& \frac{a^2}{4b}\left(1-\frac{b^2}{a^2}\right)^2\left(\phi-\phi'\right)^2.
\end{eqnarray}
Note that while ellipses become needles as $b\to0$, Eq.~\eqref{eq:d_needle} corresponds
to a different branch of solutions, and is not obtained
as a limit of the above expression. Equation~(\ref{eq:d_ellipse}) thus cannot be used to 
examine the crossover to needles.
}
\item{Metronomes are a singular limit in which $d=2b$ for small tilts
(and given by Eq.~\eqref{eq:d_needle} for larger tilts). The high density configuration
is not unique and resembles that of hard circles.
}
\end{itemize}

From these results, we observe that the general expansion for symmetric objects
has the form
\begin{equation}\label{eq:d_general}
d(\phi,\phi')=d_0+d_1(\phi^2+\phi'^2)+d_2\left|\phi-\phi'\right|^\gamma+\cdots\,.
\end{equation}
Here, $d_0$ indicates the separation at jamming, while $d_1$ indicates how 
the footprint of the particle is increased as it tilts. For needles $d_0=d_1=0$, 
while $d_1=0$ for metronomes. The correlations between orientations are due to 
the third term in the expansion, and we note that generically $\gamma=1$ when 
the intersection point is away from the line (as in needles and rectangles), 
while $\gamma=2$ when it is close to zero for small tilts (as for ellipsoids).
Lebowitz {\it et al.}~\cite{lebowitz} allow for the possibility of a general
$\gamma$, although this will only be the case for singular shapes and is not  generic. 
In any case, for $\gamma\leq 2$ we can ignore the subdominant term proportional
to $d_1$ in a saddle-point evaluation of the final integrals in Eq.~\eqref{eq:Zdef}.
We can then change variables to $w=\beta p d_2\left|\phi-\phi'\right|^\gamma$,
and simple dimensional analysis leads to 
${\cal Z} \propto{e^{-N\beta p d_0}}{(\beta p)^{-N(1+\gamma^{-1})}}$, and 
\begin{equation}\label{eq:meanL}
x_N\equiv L=-\frac{\partial\ln{\cal Z}}{\partial(\beta p)}
\approx N\left(d_0+\frac{(1+\gamma^{-1})}{\beta p}\right)\,.
\end{equation}
We can define the free space available for fluctuating particles as 
$\ell_f\approx (1+\gamma^{-1})/\beta p$, and in the remainder we shall use
this expression to switch between $\beta p$ and $\ell_f$. (Note that the 
saddle point estimate is only asymptotically exact at high pressures.)

To study orientational fluctuations and correlations, we coarse-grain 
the variables $\{\phi_i\}$ to a continuous field $\phi(n)$. From symmetry 
grounds, we expect that configurations of the coarse-grained field are 
distributed according to the Boltzmann weight of a Hamiltonian
\begin{equation}\label{eq:bH}
\beta{\cal H}[\phi]=\frac{1}{2}\int \upd n \left[J\phi^2 
+ K\left(\frac{\partial\phi}{\partial n}\right)^2+\cdots\right]\,.
\end{equation}
For the case of ellipses, with $\gamma=2$, Eqs.~(\ref{eq:Zdef},\ref{eq:d_general})
immediately
suggest $J\approx 4\beta p d_1$ and (replacing differences by derivatives)
$K\approx 2\beta p d_2$.  In the limit of $a\gg b$, Eq.~\eqref{eq:d_ellipse} 
leads to $J\approx 2\beta pb$ and $K\approx \beta p a^2/(2b)$.

It is less clear how a discrete sum of distances as in Eq.~\eqref{eq:d_needle} 
for needles goes over to the continuum form of Eq.~\eqref{eq:bH}. However, if 
we integrate (in the saddle point sense) the weight in Eq.~\eqref{eq:Zdef} over 
every other (say odd) $\phi_i$, the effective weight over the remaining (even) 
$\phi_i$s takes the form
\begin{equation}\label{eq:CGneedles}
 -\beta H_{\rm eff}\approx -\frac{1}{2}\sum_{i\ \rm{odd}}\left[ (\beta pa)^2(\phi_{i+1}-\phi_{i-1})^2+
\phi_i^2\right].
\end{equation}
In the continuum limit this then leads to $J\approx 1/2$ and 
$K\approx 2(\beta pa)^2$.
The appearance of a finite $J\approx 1/2$ is worth noting: while at the lowest 
order there is no cost to tilting a needle, the nonlinearities implicit in 
the denominator of Eq.~(\ref{eq:d_needle}) create a preference for $\phi=0$ 
after a single integration.  While for needles this is the leading 
term, in general we expect such contributions from non-linearities 
as corrections to the larger terms proportional to powers of $\beta p$.
To recover the above result for needles as $(b/a)\to 0$, we may employ
the interpolation formula 
$J\approx 2\beta p b+1/2$ for both ellipses and rectangles (from Eq.~\eqref{eq:d_rectangle}).
While we shall use this interpolation form in the remainder of the paper, we
do not expect it to hold exactly (discreteness and non-linearities are likely to
lead to a more complicated formula). 
As in needles, we get 
$K\approx 2(\beta p a)^2$ for rectangles, while in the case of metronomes, 
we expect a finite value of $K$ to result from non-linearities.  

From the Gaussian weight associated with Eq.~\eqref{eq:bH}, it is easily 
obtained that the fluctuations in angle for each object are characterized 
by a variance $\sigma^2\equiv \langle\phi^2\rangle=(4JK)^{-1/2}$. Using the 
asymptotic forms for $J$ and $K$ obtained above, this leads to
\begin{equation}\label{eq:var_rec_ell}
\sigma^2\simeq
\begin{cases}
\frac{1}{2f\sqrt{1+4\alpha f}}, &\tx{for rectangles}\\
\frac{\sqrt{\alpha}}{\sqrt{f(1+4\alpha f)}} , &\tx{for ellipses}
\end{cases},
\end{equation}
where we have introduced the dimensionless parameters - reduced pressure
$f=\beta p a$ and aspect ratio $\alpha=b/a$.
We numerically checked these results by utilizing the transfer matrix
\cite{kardar,baxter,casey}
$M(\phi,\phi')=e^{-\beta p d(\phi,\phi')}$ to evaluate the integrals in Eq.~\eqref{eq:Zdef}. 
The angular variables were discretized to obtain finite matrices, which were then
multiplied together as detailed in Ref.~\cite{kkelast}, where the transfer matrix
method was employed for computation of elastic moduli. 
Naturally in the numerical computations we used the exact form of the
separation $d(\phi,\phi')$ for rectangles and ellipses, and not the asymptotic forms given
above for the saddle point evaluations. 

%%%%%%%%%%%%%%%%%%%%%%%%%%%%%%%%%%%%%%%%%%%%%%%%%%%%%%%%%%
\begin{figure}
\onefigure[width=7.5cm]{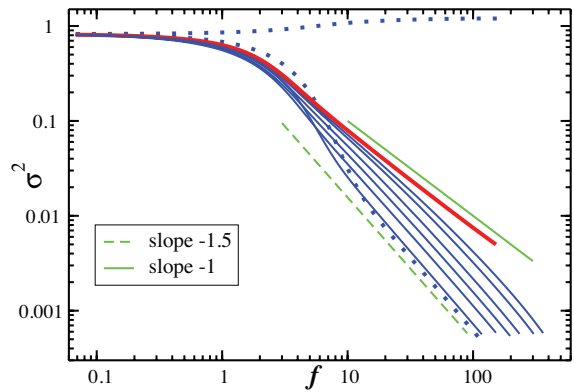}
\caption{\label{fig:varphi_rec}(Color online) Variance of angle as a function of dimensionless
pressure $f=\beta p a$ for rectangles. The rightmost (red) solid line is for needles
($\alpha=0$), while the remaining solid lines 
correspond to aspect ratios $\alpha=0.01,0.02,0.05,0.1,0.2,0.5$ (right-to-left). 
The lower dotted line demonstrates the fast decay for $\alpha=0.7$, 
while the almost horizontal dotted line corresponds to a square particle $\alpha=1$. 
Short straight segments of slopes -1 and -1.5 are included for visual comparison.
}
\end{figure}
%%%%%%%%%%%%%%%%%%%%%%%%%%%%%%%%%%%%%%%%%%%%%%%%%%%

Figure~\ref{fig:varphi_rec} depicts the variance in angular fluctuations, as a function
of reduced pressure $f$, for rectangles with several aspect ratios. 
For $f<1$, all graphs have a constant value corresponding to a uniform distribution of orientations, since in this dilute  gas regime the orientations are unconstrained. 
For larger $f$, the curves for different values of $\alpha$ initially follow the behavior
of a needle ($\alpha=0$), with $\sigma^2\propto 1/f$, but 
then depart, and eventually scale as $\sigma^2\sim 1/f^{3/2}$, which 
characterizes the variance of rectangles. The crossover occurs at larger $f$ for smaller 
aspect ratios, as predicted by 
Eq.~\eqref{eq:var_rec_ell}. The case of $\alpha=1$ is exceptional, 
because the rectangle becomes a square, with two preferred orientations 
instead of one.

%%%%%%%%%%%%%%%%%%%%%%%%%%%%%%%%%%%%%%%%%%%%%%%%%%%%%%%%%%%
\begin{figure}
\onefigure[width=7.5cm]{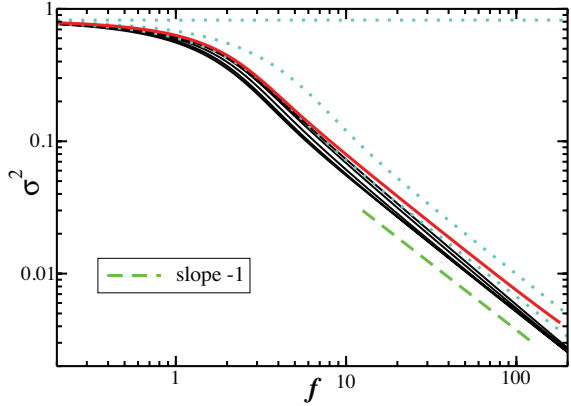}
\caption{\label{fig:varphi_ell}(Color online) Variance of angle 
as a function of dimensionless
pressure $f$ for ellipses. The rightmost (red) solid line is for needles
($\alpha=0$), while remaining solid lines correspond to 
aspect ratios $\alpha=0.01,0.02,0.05,0.1,0.2$ (right-to-left). 
Dotted lines (left-to-right) demonstrate the function for $\alpha=0.5,0.7$, 
while the horizontal dotted line correspond to a circular particle $\alpha=1$.
The short straight segment of slope -1 is included for visual comparison.
}
\end{figure}
%%%%%%%%%%%%%%%%%%%%%%%%%%%%%%%%%%%%%%%%%%%%%%%%%%%%%%%%%%%

The pressure-dependence of the  variance of the angle for ellipses
with different aspect ratios is depicted in Fig.~\ref{fig:varphi_ell}.
Again all curves coincide for $f<1$, with a value corresponding to a uniform distribution
in angles. Upon increasing $f$, the variance approaches the scaling
form $\sigma^2\sim 1/f$ predicted by Eq.~\eqref{eq:var_rec_ell}. Note that the overall trend
of the graphs moving right-to-left with increasing $\alpha$ is reversed
for $\alpha=0.5$, and eventually for $\alpha=1$ the ellipse becomes
a circle with all orientations equally probable, independent of $f$.

We also numerically examined correlations of the nematic order parameter
$s\equiv \cos 2\phi$. The  two-point correlation function 
$c(n)=\langle s(i)s(i+n)\rangle-\langle s\rangle^2$ decays from 
$\langle\cos^2(2\phi)\rangle$
for $n=0$ to 0 as $n\to\infty$.
From the Gaussian weight associated with Eq.~\eqref{eq:bH}, 
it can be verified that  
\begin{equation}\label{eq:GaussCorr}
c(n)=2\exp\left(-\frac{2}{\sqrt{JK}}\right)\sinh^2\left(\frac{1}{\sqrt{JK}}{\rm e}^{-\sqrt{{J}/{K}}n}\right).
\end{equation}
Specializing further to the limit $f\gg 1$, required for the validity of the Gaussian approximation,
we find $c(n)\propto {\rm e}^{-n/\xi}$, with $\xi^2=K/(4J)$, and in particular
\begin{equation}\label{eq:xif}
\xi^2\simeq
\begin{cases}\frac{f^2}{1+4\alpha f}, &\tx{for rectangles}\\
\frac{f}{4\alpha(1+4\alpha f)} , &\tx{for ellipses}
\end{cases}.
\end{equation}
Note that the correlation length diverges as $f\to\infty$ for rectangles, 
but asymptotes to a constant for ellipses.

\begin{figure}
\onefigure[width=7.5cm]{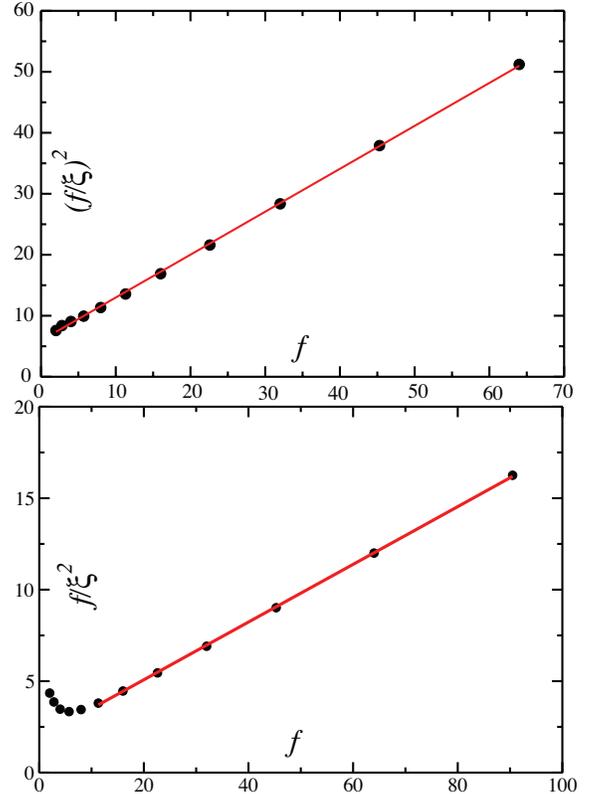}
\caption{\label{fig:xif}(Color online) Numerical verification of Eqs.~\eqref{eq:xif} 
for the asymptotic relation between the
correlation length $\xi$ and the dimensionless pressure $f$.
The top panel is for rectangles
with aspect ratio $\alpha=0.05$, and the bottom panel is for ellipses
with aspect ratio $\alpha=0.1$.
}
\end{figure}

Correlation functions were obtained numerically, by appropriate transfer matrix manipulations,
for a variety of rectangles and ellipses with different aspect ratios, and confirm the
above predictions. In particular, we verified Eqs.~\eqref{eq:xif} for a moderate value of the aspect ratio
($\alpha=0.05$ for rectangles, and $\alpha=0.1$ for ellipses)
chosen such that the crossovers in $f$ are clearly seen. 
In the plots of Fig.~\ref{fig:xif},  $\xi$ is rescaled in such a way that Eq.~\eqref{eq:xif}
is verified if the curve is a straight line. Indeed for large values
of $f$, where the Gaussian approximation  is valid: the dependence is a straight line.
The slopes and the intercept
points of the lines are of the order predicted by Eq.~\eqref{eq:xif} but do not
agree quantitatively, since the approximate nature of our derivation
does not ensure accuracy of the coefficients in Eq.~\eqref{eq:xif}.

%%%%%%%%%%%%%%%%%%%%%%%%%%%%%%%%%%%%%%%%%%%%%%%%%%%%%%%%%%RECTANGLE
%%%%%%%%%%%%%%%%%%%%%%%%%%%%%%%%%%%%%%%%%%%%%%%%%%%%%%%%%%

The most notable feature of our results is the possible divergence of the 
dimensionless correlation length $\xi$. This diverging correlation length 
cannot be associated with the onset of long-range
order (as in say the 1D Ising model), as there is a finite nematic order
$\langle \cos(2\phi)\rangle\approx \exp\left(-1/\sqrt{JK}\right)$ at any density. 
As such, the onset of jamming~\cite{jamming} is a better descriptor: As pressure/density increase,
and the free space $\ell_f$ vanishes, a larger collection of particles must be moved in concert.
This number (correlation `length') diverges for rectangles but saturates for ellipses 
(see Eq.~\eqref{eq:xif} for $f\to\infty$). 
The differences between the two cases is that 
for rectangles the point of contact is remote from the center, almost at a distance $a$.
Thus, given a typical spacing of $\ell_f$, near-neighbor angular shifts
are small and vanish as $\delta\phi\sim \ell_f/a$.
By contrast, for ellipses the point of contact goes to zero continuously, 
roughly as $\sqrt{R\ell_f}$, where $R$ is the radius of curvature at the point of contact.
This results in larger near-neighbor angular
shifts of $\delta\phi\sim\sqrt{\ell_f/R}$.
At the global scale, fluctuations in angle are limited (due to the broken nematic symmetry imposed
by the line of movement) and given by Eq.~\eqref{eq:var_rec_ell}, according to which
$\sigma$ is proportional to $\ell_f^{1/2}$, $\ell_f^{3/4}$ and $\ell_f^{1/2}$ for needles, rectangles,
and ellipses respectively. 
As $\ell_f\to0$, the local (near-neighbor) fluctuations are asymptotically smaller than global ones
for rectangles (and needles) but not for ellipses. The fluctuations must thus build over a large
correlation number in the former cases. 

This work was supported by the Israel Science Foundation Grant 
No. 99/08 (Y.K.) and by the National Science Foundation Grant 
No. DMR-08-03315 (M.K.). Part of this work was carried out at the 
Kavli Institute for Theoretical Physics, with support from  NSF 
Grant No. PHY05-51164.

 \end{document}